\documentclass[12pt]{article}

\usepackage{graphicx}

\begin{document}

\begin{titlepage}
\begin{flushright}

NTUA--99--2001


\end{flushright}

\begin{centering}
\vspace{.41in}
{\large {\bf Inflation Induced by Vacuum Energy and Graceful Exit
from it.} }\\

\vspace{.4in}

 {\bf E.~Papantonopoulos}$^{a}$ and {\bf I.~Pappa} $^{b}$ \\
\vspace{.2in}

 National Technical University of Athens, Physics
Department, Zografou Campus, GR 157 80, Athens, Greece. \\

\vspace{1.0in}

{\bf{ Abstract}}
 \\
\end{centering}

 \vspace{.1in}

Motivated by brane cosmology we solve the Einstein equations with
a time dependent cosmological constant. Assuming  that at an early
epoch the vacuum energy scales as $1/logt $, we show that the
universe passes from a fast growing phase (inflation) to an
expanding phase in a natural way.

\vspace{2.5in}
\begin{flushleft}

$^{a}$ e-mail address:lpapa@central.ntua.gr \\$ ^{b}$ e-mail
address:gpappa@central.ntua.gr

\end{flushleft}

\end{titlepage}

\section{Introduction}

The cosmological constant remains one of the biggest mysteries of
our universe. Recent observations of type Ia Super Novae suggest
that the expansion of the universe is accelerating now
\cite{Accel}. This acceleration could be explained by a positive
time dependent cosmological constant with $\Omega_{\Lambda}\sim$
0.7. This suggest that the cosmological constant is almost the
same order of magnitude as the present mass density of the
universe.

If we accept that the cosmological constant is a time dependent
quantity, to explain its present value it should be decaying with
time. If we parametrize it as $\Lambda=t^{-n}$ then considerations
on the age of the universe constrain the value of n. Adopting the
commonly used value of Hubble parameter $H_{0}=73\pm 10 km s^{-1}
Mpc^{-1} $ \cite{Hubb}, the value of n is fixed to n=2. At the
early stages of the cosmological evolution we expect the
cosmological constant to be large but we do not know anything
about its nature and how it scales with time.

Recently the idea that we are living on a D3-brane was proposed.
In this brane universe scenario, the Standard Model gauge bosons
as well as charged matter arise as fluctuations of the D-branes.
The universe is living on a collection of coincident branes, while
gravity and other universal interactions is living in the bulk
space \cite{Pol}. For all such theories, an essential issue
concerns the cosmological evolution of our universe. In the
literature, there are a lot of cosmological models associated to
brane universe \cite{Cosm}-\cite{Keh}. In a class of these models,
specially when there is a movement of the brane in the bulk
\cite{Cha}-\cite{Keh}, there is always a time dependent
cosmological constant which is induced on the brane
\cite{Papa}-\cite{Pappa}.

In \cite{Pappa} we studied the motion of the brane universe in a
type 0 string background. We calculated the energy density which
is induced on the brane because of its movement in the background.
What we found is that there is an inflationary phase of the brane
universe which is followed by another phase in which the scale
factor $\alpha$ is slowing down and the energy density is
dominated by a term proportional to $
\frac{1}{(\log(\alpha))^{4}}$. We attributed this behaviour to
"mirage" matter which is induced on the brane because of the
specific type-0 string background.

In this work we will study this effect in details. Our brane
cosmological picture tells us that an effective energy density is
induced which controls the cosmological evolution of the brane
universe. This effective energy density plays the $r\hat{o}le$ of
a time dependent cosmological constant. Motivated by this result,
we will examine the effect of a time dependent cosmological
constant to the usual Friedman-Robertson-Walker Universe. We will
consider the Einstein equations with a cosmological constant
$\Lambda(t)=\frac{1}{logt}$, and concentrate on the exact
solutions which exhibit inflationary behaviour and graceful exit
from inflation.

What we found is that the universe at its early stages of
evolution has two phases. One is of rapid expansion (inflation),
with large cosmological constant and an energy density which
decreases and reaches the critical energy density at the end of
inflation. This phase is followed by a phase of slow expansion
during which the cosmological constant remains large. The
transition from one phase to the other is continuous and smooth,
without having any singularity.

 Our work is organized as follows. In section two, we
discuss the general framework of a brane moving in the background
of other branes. In section three we solve the Einstein equations
with a time dependent cosmological constant, and finally in
section four we discuss our results.

 \section{Brane cosmology}

 There are various ways to do cosmology on the brane. The more
 conventional way is to assume that the brane is static comparable
 to the bulk background, and is situated in a particular point
 $r_{0}$ of the radial coordinate. Then matter on the brane or in
 the bulk can lead to a cosmological evolution of the brane universe.
 Another approach is to consider a brane moving in the background
 of other branes. Then one can show that even if there is no matter
 on the brane, the brane universe undergoes a cosmological evolution as
 a result of its movement \cite{Cha}-\cite{Youm}.

 The simplest way to follow such a movement is to assume that the
 brane as it moves in the bulk does not back react with the
 geometry. In \cite{Papa} and \cite{Pappa} we considered a probe
 brane moving on a geodesic
 in a generic static, spherically symmetric background \cite{Keh}.
 The metric of a D3-brane was parametrized as
 \begin{equation}\label{in.met}
ds^{2}_{10}=g_{00}(r)dt^{2}+g(r)(d\vec{x})^{2}+
  g_{rr}(r)dr^{2}+g_{S}(r)d\Omega_{5}
\end{equation}
 In the background there is also a dilaton field $\Phi$ and a $RR$
 background~$C(r)=C_{0...3}(r)$ with a self-dual field strength.
 The dynamics on the brane will be governed by the
 Dirac-Born-Infeld action  given by

\begin{eqnarray}\label{B.I. action}
  S&=&T_{3}~\int~d^{4} \xi
  e^{-\Phi}\sqrt{-det(\hat{G}_{\alpha\beta}+(2\pi\alpha')F_{\alpha\beta}-
  B_{\alpha\beta})}
   \nonumber \\&&
  +T_{3}~\int~d^{4} \xi\hat{C}_{4}+anomaly~terms
\end{eqnarray}
 One can show that the induced metric on the brane is \cite{Keh}
 \begin{equation}\label{fin.ind.metric}
d\hat{s}^{2}=-d\eta^{2}+g(r(\eta))(d\vec{x})^{2}
\end{equation}
 with $\eta$ the cosmic time which is defined in terms of the background
 fields by
\begin{equation}\label{cosmic}
 d\eta=\frac{|g_{00}|g^{\frac{3}{2}}e^{-\Phi}}{|C+E|}dt
\end{equation}
The induced metric on the brane, is the standard form of a flat
expanding universe. For this metric we can write the effective
Einstein equations on the brane,
\begin{equation}\label{Einst1}
  R_{\mu \nu}-\frac{1}{2} g_{\mu \nu} R =8\pi G (T_{\mu \nu})_{eff}
\end{equation}
where $(T_{\mu\nu})_{eff}$ is the effective energy momentum tensor
which is induced on the brane. We have assumed that our brane is
light and there is no back-reaction with the bulk. We expect
$(T_{\mu \nu})_{eff}$ to be a function of the quantities of the
bulk. If we now assume the usual form of a perfect fluid for the
effective energy momentum tensor, we get from  (\ref{Einst1})
\begin{equation} \label{r-tt}
 8 \pi G \rho +\Lambda= \frac{3}{4} g^{-2}\dot{g}^{2}
 \end{equation}
We can define a $\rho_{eff}$ from the relation
 \begin{equation}\label{ident}
  8 \pi G \rho +\Lambda= 8 \pi G_{N} \rho_{eff}
 \end{equation}
Using equation (\ref{cosmic}) we get
\begin{equation} \label{gprim}
 \dot{g}=g^{\prime}\Big{[} \frac {|g_{00}|}{g_{rr}}
 -\frac {g_{00}^{2}}{g_{s} g_{rr}} \Big{(}
 \frac{ g^{3}g_{s}e^{-2\Phi}+\ell^{2} }
 {(C+E)^{2}} \Big{)}
 \Big{]} ^{\frac{1}{2}}
  \frac{|C+E|}
  {|g_{00}|g^{\frac{3}{2}} e^{-\Phi}}
\end{equation}
where prime denotes differentiation with respect to $r$.
 To derive an analogue of the four dimensional Friedman equations
 for the expanding four dimensional universe on the probe
 D3-brane, we define the scale factor as $g=\alpha^{2}$ and then equation
(\ref{ident}) with the use of  (\ref{r-tt}) and (\ref{gprim})
becomes
\begin{equation}\label{dens}
 \frac{8\pi}{3}G_{N}\rho_{eff}= (\frac
{\dot{\alpha}}{\alpha})^{2}=
\frac{(C+E)^{2}g_{S}e^{2\Phi}-|g_{00}|(g_{S}g^{3}+\ell^{2}e^{2\Phi})}
{4|g_{00}|g_{rr}g_{S}g^{3}}(\frac{g'}{g})^{2}
\end{equation}

Therefore the motion of a D3-brane on a general spherically
symmetric background had induced on the brane a matter density. As
it is obvious from the above relation, the specific form of the
background will determine the cosmological evolution on the brane.
If we go now to a particular background namely of a type 0 string,
we get for $\rho_{eff}$ \cite{Pappa}
\begin{eqnarray}\label{rhonimah}
\frac{8\pi}{3}\rho_{eff}& =&
 \Big{[} \Big{(}1-\frac{1}{Q\alpha^{4}}Ei[log2Q + 4log\alpha] +
\frac{E} {2\alpha^{4}} \Big{)}^{2} \nonumber \\ && -
\frac{1}{4}\Big{(}1-\frac{1}{2(log2Q + 4log\alpha)}\Big{)}^{4}
\Big{ ]}\Big{(}1-\frac{1}{2(log2Q +
4log\alpha)}\Big{)}^{-4}\nonumber \\ && \Big{(}1-\frac{9}{2(log2Q
+ 4log\alpha)}\Big{)}^{-1} \nonumber
\\&& \Big{(}1+\frac{1}{(log2Q +
4log\alpha)^{2}}\frac{1}{\Big{(}1-\frac{1}{2(log2Q +
4log\alpha)}\Big{)}}\Big{)}^{2}
\end{eqnarray}
As we can see from the above relation, the brane universe has an
inflationary phase corresponding to the constant term. As the
scale factor $\alpha$ evolves, the term
$\frac{1}{(\log\alpha)^{4}}$ starts to contribute to $\rho_{eff}$
having the effect of slowing down the exponential growth.
Therefore on the brane we have an inflationary phase which is
followed by a less rapid expansion.

The energy density induced on the brane is time dependent, because
the brane moves in the bulk. This time dependent energy density,
effectively acts, as a time dependent cosmological constant. One
should be careful though of introducing a time dependent
cosmological constant in the usual four-dimensional Einstein
equations. Consider the four-dimensional Einstein equations with a
time dependent cosmological constant $\Lambda$
\begin{equation}\label{einst}
G_{\mu\nu}=8\pi G T_{\mu\nu}-\Lambda g_{\mu\nu}
\end{equation}
where $G_{\mu\nu}=R_{\mu\nu}-Rg_{\mu\nu}/2$ is the Einstein tensor
and $T_{\mu\nu}$ is the energy momentum tensor. Taking the
covariant divergence of equation (\ref{einst}) and assuming the
conservation law $\nabla^{\nu}T_{\mu\nu}$ it follows that
$\Lambda=constant$. Therefore a time dependent cosmological
constant violates energy conservation in general relativity.

After the recent astrophysical observations, we know that the
dominant contribution to the energy density comes from the
invisible or dark matter. Then one can define an effective energy
momentum tensor by
\begin{equation}\label{effen}
\tilde{T}_{\mu\nu}\equiv T_{\mu\nu}-\frac{\Lambda}{8\pi
G}g_{\mu\nu}
\end{equation}
where $T_{\mu\nu}$ is the contribution of the normal matter and
 $ \frac{\Lambda} {8\pi G} g_{\mu\nu} $ is the dark matter
 component of the energy momentum
tensor. Then these is no any compelling season why
$\tilde{T}_{\mu\nu}$ should not be conserved.

A time dependent cosmological constant, except its usefulness in
the standard cosmology to parametrize the unknown nature of the
dark matter \cite{cocos}, it is a very useful concept in brane
cosmology. The brane as it moves, exchanges energy with the bulk
and the presence of a time dependent cosmological constant is
necessary to maintain the energy conservation. An observer in our
brane-universe has as a dynamical variable the scale factor
$\alpha$. Therefore all the quantities in his world are expressed
as functions of $\alpha$ and he writes the Einstein equations as
relations (\ref{Einst1}). For this observer the energy momentum
tensor $(T_{\mu\nu})_{eff}$ is a well behaved conserved quantity.
An observer now in the bulk can write the Einstein equations as
the relations (\ref{einst}). For him the energy momentum tensor
$T_{\mu\nu}$ and the cosmological constant $\Lambda$ are functions
of the radial coordinate r. As the brane moves along the radial
coordinate, there is a constant flow of energy between the bulk
and the brane.  The two pictures will coincide eventually, when
the brane is in large radial distance and does not feel the
gravitational field of the other branes.

One can see more clearly the transfer of energy between the bulk
and the brane in the case where the brane is "fat" and back-reacts
with the background. Assume that the bulk is D-dimensional and
contains a brane (domain wall). Demanding that the metric be
continuous everywhere and that the derivatives of the metric be
continuous everywhere except on the brane, one can derive, using
the Israel equations, the equations \cite{Cha}
\begin{equation}
\bar{\nabla }_{\mu}t^{\mu \nu}=-\Big{\{} \bar{\nabla }_{\mu}K^{\mu
\nu} -h^{\mu\nu} \bar{\nabla }_{\mu}K  \Big{\}}.
\end{equation}
where $t_{\mu\nu}$ is the energy momentum tensor on the brane
$K_{\mu\nu}$ is the extrinsic curvature and $h_{\mu\nu}$ is the
induced metric on the brane. This equation shows that the energy
on the brane is not conserved. Using Codacci's equation we get
\begin{equation}
\bar{\nabla }_{\mu}t^{\mu \nu}=-\Big{\{} h^{\nu\mu} T_{\mu  p}
n^{p} \Big{\}}
\end{equation}
where $T_{\mu\nu}$ is the energy momentum tensor of the bulk. This
equation describes conservation of energy as it flows from the
bulk to the brane and vice versa.

\section{Einstein equations with a time dependent cosmological
constant}

Motivated by our brane universe picture, we will assume that at
early stages of cosmological evolution, there is a time dependent
effective cosmological constant, which scales as
$\Lambda(t)=\frac{1}{logt} $, and study its effect to the usual
Friedman-Robertson-Walker universe.
 We will work in a homogeneous and isotropic Universe
with the energy momentum tensor having the form of a perfect fluid
with pressure p and energy density $\rho$. In accordance to
equation (\ref{effen}) we define an effective pressure $
\tilde{p}$ and an effective energy density $ \tilde{\rho}$ from
\begin{eqnarray}\label{efpre}
\tilde{p} &\equiv& p-\frac{\Lambda}{8\pi G} \nonumber \\
 \tilde{\rho} &\equiv&
\rho-\frac{\Lambda}{8\pi G}
\end{eqnarray}
The field equations (\ref{einst}) and the energy conservation give
\begin{equation}\label{sceins}
\dot{\alpha}^{2}=\frac{8\pi
G}{3}\rho\alpha^{2}+\frac{\Lambda}{3}\alpha^{2}-k
\end{equation}
\begin{equation}\label{consen}
\frac{d}{d\alpha}(\rho\alpha^{3\gamma})=-\Big{(}\frac{\alpha^{3\gamma}}{8\pi
G}\Big{)} \frac{d\Lambda}{d\alpha}
\end{equation}
In deriving equation (\ref{consen}) we have taken as an equation
of state
\begin{equation}
p=(\gamma-1)\rho
\end{equation}
with $\gamma=constant$. We get the dust-dominated universe for
$\gamma=1$ and the radiation-dominated universe for $\gamma=4/3$.
Differentiating equation (\ref{sceins}) and using (\ref{consen})
we get
\begin{equation}\label{crit}
\ddot{\alpha}=\frac{8 \pi G}{3}\Big{(}1-\frac{3\gamma}{2}\Big{)}
\rho\alpha+\frac{\Lambda}{3}\alpha
\end{equation}
The above equation shows that if $\gamma>2/3$ a positive $\rho$
acts to decelerate the expansion, while if $\gamma<2/3$ the
density accelerates the expansion. Combining equations
(\ref{sceins}) and (\ref{crit}) we get
\begin{equation}\label{common}
\frac{\ddot{\alpha}}{\alpha}=\Big{(}1-\frac{3\gamma}{2}\Big{)}
\Big{(}\frac{\dot{\alpha}^{2}}{\alpha^{2}
}+\frac{k}{\alpha^{2}}\Big{)}+\frac{\gamma}{2}\Lambda
\end{equation}
This is the equation that governs the behavior of the scale factor
in the presence of a cosmological term $\Lambda$. Introducing
$H=\frac{\dot{\alpha}}{\alpha}$ we obtain from (\ref{common})
\begin{equation}\label{hcom}
\dot{H}=-\frac{3\gamma}{2}H^{2}+\frac{\gamma}{2}\Lambda+
\Big{(}1-\frac{3\gamma}{2}\Big{)}\frac{k}{\alpha^{2}}
\end{equation}
We will work with spatially flat universes and in this case
because k=0 the last term of equation (\ref{hcom}) drops off
leaving a special case of Riccati's equation \cite{over}. Defining
a new variable $ x $ from the relation
\begin{equation}\label{hcon}
H=\Big{(}\frac{2}{3\gamma}\Big{)}\frac{\dot{x}}{x}
\end{equation}
equation (\ref{hcom}) becomes
\begin{equation}\label{xeqa}
\frac{\ddot{x}}{x}-\frac{3\gamma^{2}}{4}\Lambda=0
\end{equation}
We take $\Lambda(t)=\frac{1}{logt}$ and to set a time scale, we
know that the inflationary period of the universe is well before
the nucleosynthesis. Therefore our assumption is valid until a
cutoff time $t_{c}$, the time before the nucleosynthesis starts,
\begin{equation}\label{lamcon}
\Lambda=\left\{\begin{array}{ccc} \frac{ A }{logt}& when& t<t_{c}
\\
\Lambda_{c}&when& t>t_{c}
\end{array}\right.
\end{equation}
where A is a constant. This choice of $t_{c}$ is consistent with
our brane world picture. We expect that this time dependent
effective cosmological constant will receive various contributions
from other interactions, mainly gauge interactions, at later
cosmological evolution stages. Substituting $\Lambda$ of
(\ref{lamcon}) in the equation (\ref{xeqa}) we get
\begin{equation}\label{loeq}
(logt)\ddot{x}-ax=0
\end{equation}
where $a=\frac{3}{4} \gamma^{2}A$. We go again to a new variable
$z$, defined by $t=exp(-b z^{q})$ where $b$ and q are constants.
Then equation (\ref{loeq}) becomes
\begin{equation}\label{maineq}
z^{2-q} \frac{d^{2}x}{d z^{2}}+\Big{(}(1-q)z^{1-q}+b q z
\Big{)}\frac{d x}{d z}+a b q^{2} x e^{-2b z^{q}}=0
\end{equation}
We are looking for solutions of the above equation for various
values of the parameters q, $a$, and b. To simplify our solutions
we fix q=1. Before we discuss in detail the solutions, let us
assume that we have a solution x(z) of the equation
(\ref{maineq}). Then using the transformation $t=e^{-b z}$ we get
x(t) and from (\ref{hcon}) the scale factor is
\begin{equation}\label{scfac}
\alpha(t)=[x(t)]^{2/3\gamma}
\end{equation}

To solve equation (\ref{maineq}) we first consider the case where
b =1. Then $t=e^{- z}$ and equation (\ref{maineq}) becomes
\begin{equation}\label{eq1}
 \frac{d^{2}x}{d z^{2}}+ \frac{d x}{d z}+a \frac{e^{-2 z}}{z} x=0
 \end{equation}
 This equation cannot be solved analytically. Nevertheless for
 small values of z equation (\ref{eq1}) can be approximated by the equation
 \begin{equation}\label{eq11}
 \frac{d^{2}x}{d z^{2}}+ \frac{d x}{d z}+a \frac{x}{z}=0
 \end{equation}
 This equation can give exact solutions for various values of $a$.
If $ a =\pm 1$ we get the following solutions
\begin{equation}\label{aeq11}
\begin{array}{ccc}
 x(z)=c_{1} e^{-z} z +c_{2}e^{-z} z \Gamma(-1,-z)  & when& a=1
\\
 x(z)=z c_{1}+c_{2}MeijerG &when& a=-1
\end{array}
\end{equation}
Where $c_{1}$ and $c _{2}$ are constants to be fixed by initial
conditions, end MeijerG is the Meijer G function
$G^{20}_{10}\Big{(}z\mid^{2}_{0,1}\Big{)}$. Equation (\ref{eq1})
for large values of $z$ can be approximated by the equation
\begin{equation}\label{eq12}
\frac{d^{2}x}{d z^{2}}+ \frac{d x}{d z}+a e^{-2z} x=0
\end{equation}
This equation can be solved analytically and for $a=\pm 1 $ we get
two solutions
\begin{equation}\label{ffsol}
\begin{array}{ccc}
 x(z)=c_{1}cos( e^{-z}) +c_{2} sin(e^{-z}) & when& a=1
\\
x(z)=c_{1}cosh( e^{-z}) +ic_{2} sinh(e^{-z})&when& a=-1
\end{array}
\end{equation}

We will discuss first the case of $a=1$. Inserting the solutions
for $a=1$ in (\ref{scfac}) we get
\begin{equation}\label{sol1}
\begin{array}{ccc}
\alpha(t)=c_{1}^{\prime} t^{2/3\gamma}(-log t)^{2/3\gamma} &
small& t
\\
\alpha(t)=c_{1}^{\prime}(cost) ^{2/3\gamma} & large& t
\end{array}
\end{equation}
We have absorbed the factor $2/3\gamma$ into the definition of
$c_{1}^{\prime}$ and have put $c_{2}=0$. This solution does not
have an inflationary phase and therefore we do not discuss it any
further. For $ a=-1$ using (\ref{scfac}) we get
\begin{equation}\label{sol2}
\begin{array}{ccc}
\alpha(t)=c_{1}^{\prime}(-log t)^{2/3\gamma} & large& t
\\
\alpha(t)=\frac{c_{1}^{\prime}}{2}(e^{t}+e^{-t})^{2/3\gamma}  &
small & t
\end{array}
\end{equation}
where again we have absorbed the factor $2/3\gamma$ into the
definition of $c_{1}^{\prime}$ and have put $c_{2}=0$. This is an
interesting cosmological solution. The universe in an early epoch
has two phases: one is rapid expansion (inflation) with  positive
cosmological constant $\Lambda$ and a slow expansion phase also
with positive cosmological constant. The crucial question is how
these two phases are connected. We performed a numerical
investigation of equation (\ref{eq1}) for $a=-1$ and for a wide
range of values of $z$. What we found is that the two phases are
connected in a smooth way in the sense that the scale factor
$\alpha(t)$ is passing from one phase to the other without having
any singularity.

Another interesting question is for how long the inflation lasts.
We can answer this question calculating the number of e-foldings.
The number of e-foldings is given by
\begin{equation}
N=\int^{t_{f}}_{t_{i}} H(t) dt
\end{equation}
where $t_{i}$ is the time when inflation starts and $ t_{f}$ the
time it ends. Using equation (\ref{hcon}) and the solution for
$x(z)$ from (\ref{ffsol}) we get
\begin{equation}
N=\frac{2}{3\gamma}ln\frac{cosht_{f}}{cosht_{i}}
\end{equation}
Demanding $N > 60$ we get
\begin{equation}
cosht_{f}>cosht_{i} e^{90\gamma}
\end{equation}
To satisfy the above equation for a very small time $t_{i}$,
$\gamma$ should be positive and very small. This value of $\gamma$
is consistent with the "inflationary zone" $\gamma<\frac{2}{3}$
and the expectation that during inflation, the pressure $p$ is
negative.

 We can also find how the energy density varies during the two phases.
Using (\ref{sol2}), the equation for the energy density
(\ref{consen}) becomes for the inflationary phase
\begin{equation}
\dot{\rho}+2\frac{e^{t}-e^{-t}}{e^{t}+e^{-t}}\rho+\frac{4}{24\pi G
\gamma^{2}} \frac{1}{\Big{(}t(logt)^{2}\Big{)}} =0
\end{equation}
Note that the energy density does not depend on the initial
conditions of the scale factor but only on $\gamma$, and the
initial value of $\rho$. A numerical analysis of the above
equation shows that as the universe expands the energy decreases.
To solve the flatness problem, we expect the energy density $\rho$
at the end of inflation to reach $\rho_{c}$ which is given by
\begin{equation}
\rho_{c}=\frac{3H^{2}}{8\pi G}
\end{equation}
Ii our case if we define $\Omega =\frac{\rho}{\rho_{c}}$ and use
(\ref{sceins}) we get for k=0
\begin{equation}
|\Omega(t)-1|=\frac{\Lambda}{3H^{2}}
\end{equation}
Then using (\ref{hcon}) with the solution $x(z)$ from
(\ref{ffsol}) we get
\begin{equation}
|\Omega(t)-1|=-\frac{(cotht)^{2}}{logt}
\end{equation}
The right hand side of this relation tends to zero for "large"
times where our approximation is still valid and the inflation
terminates. Therefore $\rho\longrightarrow \rho_{c}$ as expected.

For the slow expansion phase, the energy density is given by the
equation
\begin{equation}
\dot{\rho}+2\frac{\rho}{t(logt)}+\frac{4}{24\pi G \gamma^{2}}
\frac{1}{\Big{(}t(logt)^{2}\Big{)}} =0
\end{equation}
We expect that during this phase the reheating starts. The vacuum
energy remains positive and is increasing providing the energy for
the reheating.

 The second class of solutions are generated by taking
$b=-1$. These solutions are similar to $b=1$ case, but we are
giving them for the completeness of our discussion. In this second
case the equation (\ref{maineq}) for small $z$ can be approximated
by
\begin{equation}\label{eq21}
 \frac{d^{2}x}{d z^{2}}- \frac{d x}{d z}+a \frac{x}{z}=0
 \end{equation}
 giving the solutions
\begin{equation}
\begin{array}{ccc}
 x(z)=c_{1}z +c_{2} MeijerG & when& a=1
\\
x(z)=c_{1} e^{z}z +c_{2} e^{z}z \Gamma(-1,z) &when& a=-1
\end{array}
\end{equation}
where MeijerG is the Meijer G function
$G^{20}_{10}\Big{(}-z\mid^{2}_{0,1}\Big{)}$. For large $z$, using
an equation similar to (\ref{eq12}) we get
\begin{equation}\label{aq22}
\begin{array}{ccc}
 x(z)=c_{1} e^{-ie^{z}} -\frac{1}{2}c_{2} ie^{ie^{z}} & when& a=1
\\
x(z)=c_{1} e^{e^{z}} -\frac{1}{2}c_{2} ie^{-e^{z}} & when& a=-1
\end{array}
\end{equation}
For $a=1$ we can not get any meaningful solution because $x(z)$
from equation (\ref{aq22}) is imaginary. For $a=-1$ the scale
factor is
\begin{equation}\label{finsol}
\begin{array}{ccc}
\alpha(t)=c_{1}^{\prime} (t log t)^{2/3\gamma} & small& t
\\
\alpha(t)=c_{1}^{\prime}e^{\frac{2t}{3\gamma}}  & large& t
\end{array}
\end{equation}
Again here we have an interesting cosmological solution with two
phases one of rapid expansion and the other of slow growth
connected in a smooth way. Note that inflation occurs at late
cosmological times.

\section{Discussion}

We have studied the Einstein equations with a time dependent
cosmological constant. What we have found is that when
$\Lambda(t)=\frac{1}{logt}$ there are exact solutions of the
Einstein equations which exhibit two distinct phases of the
cosmological evolution. The first phase is an inflationary phase
and the second is a slow expanding phase. These two phases are
connected by a smooth way without any singularity. Such
cosmological evolution can be realized in a brane universe
scenario. In a situation where a probe brane is moving in the
gravitational field which is produced by other branes, a time
dependent cosmological constant will be induced on the brane,
which drives the inflation and at later times terminates it.

Our model resembles the old DeSitter model. In that model the
vacuum energy is constant. This constant vacuum energy can drive
the universe to a flat geometry but this evolution if it is
allowed to continue will eventually produce an empty universe.

In our model the vacuum energy is time dependent and has a large
value comparable to its present value. Our aim was to show that
the inflationary phase terminates and the scale factor follows a
slow expansion in a continuous way. For this reason we cut-off the
vacuum energy for some time before the nucleosynthesis. We do not
know how the vacuum energy will evolve for later times and reach
its present value. This is consistent with our brane-world
scenario, which is the motivation of this work. We expect that the
vacuum energy will receive various contributions from other
interactions, mainly gauge interactions at later cosmological
times.

We had studied some very basic tests of our model, like the
duration of inflation and how the energy density varies with time.
A more detailed phenomenological analysis is needed and problems
like the reheating of the universe should be answered. Finally a
more general form of the cosmological constant $\Lambda(t)=\Big{(}
\frac{1}{log \alpha(t)} \Big{)}^{q}$ should be investigated
\cite{mousi}.

\section*{Acknowledgement}
We would like to thank C. Bachas, A. Kehagias and E. Kiritsis for
valuable discussions. We thank specially N. Mavromatos for his
constructive remarks. Partially supported by NTUA program
Archimedes.


\begin{thebibliography}{99}


\bibitem{Accel}
S. Perlmutter et al., Nature {\bf{391}}, 51 (1998); A. G. Riess et
al., Astron. J. {\bf{116}}, 1009 (1998); P. M. Garnavich et al.,
Astrophys. J. {\bf{509}}, 74 (1998); S. Permutter et al.,
Astrophys. J. {\bf{530}}, 17 (2000).

\bibitem{Hubb}
W. L. Freedman, Los Alamos report [astro-ph/9612024] (1996).

\bibitem{cocos}
E. B. Gliner, Sov. Phys. JETP {\bf 22}, 378 (1966); Y. B.
Zel'dovich, Sov. Phys. Usp. {\bf 11}, 381 (1968); A. D. Linde,
JETP letters {\bf 19}, 183 (1974); A. M. Polyakov, Sov. Phys. Ups.
{\bf 25}, 187 (1982); S. L. Adler, Rev. Mod. Phys. {\bf 54} 729
(1982).

\bibitem{over}
J. M. Overduin and F. I. Cooperstock, Phys. Rev. {\bf D58} 043506
(1998).

\bibitem{Pol}
J.Polchinski, {\it Dirichlet branes and Ramond-Ramond charges,
Phys. Rev. Lett.} {\bf{75}} {\it (1995) 4724} [hep-th/9510017]


\bibitem{Cosm}
N.Kaloper and A.Linde, {\it Inflation and large internal
dimensions, Phys. Rev.}  {\bf{D 59}} {\it (1999) 101303}
[hep-th/9811141];

N.Arkani-Hamed, S.Dimopoulos, N.Kaloper and J.March-Russell, {\it
Rapid asymmetric inflation and early cosmology in theories with
submillimeter dimensions,} [hep-ph/9903224];

N.Arkani-Hamed, S.Dimopoulos, G.Dvali and N.Kaloper, {\it
Infinitely large new dimensions,} hep-th/9907209;

N.Kaloper. {\it Bent domain walls as brane-worlds, Phys. Rev.}
{\bf{D60}} {\it (1999) 123506} [hep-th/9905210].

P.Kanti, I.I.Kogan, K.A.Olive and M.Pospelov, {\it Cosmological
3-Brane Solution, Phys. Lett.} {\bf{B468}} {\it (1999) 31}
[hep-ph/9909481]

\bibitem{Stat}
P.Binetruy, C.Deffayet and D.Langlois, {\it Nonconventional
cosmology from a brane universe,} [hep-th/9905012].

G.Dvali and S.H.H.Tye, {\it Brane inflation, Phys. Lett.}
{\bf{B450}} {\it (1999) 72} [hep-ph/9812483];

E.E.Flanagan, S.H.H.Tye and I.Wasserman, {\it A cosmology of the
brane world,} [hep-ph/9909373].

H.B.Kim and H.D.Kim, {\it Inflation and Gauge Hierarchy in
Randall-Sundrum Compactification,} [hep-th/9909053]

C.Csaki, M.Graesser, C.Kolda and J.Terning, {\it Cosmology of One
Extra Dimension with Localized Gravity, Phys. Lett.} {\bf{B462}}
{\it (1999) 34} [hep-ph/9906513]

C.Csaki, M.Graesser, L.Randall and J.Terning, {\it Cosmology of
Brane Models with Radion Stabilization,} [hep-ph/9911406]

S. Nojiri, O. Obregon and S. D. Odintsov, {\it (Non)-Singular
Brane-World Induced by Quantum Effects in D5 Dilatonic Gravity,
Phys. Rev.}, {\bf D62} {\it (2000) 104003}, [hep-th/0101003].

\bibitem{Cha}
H.A.Chamblin and H.S.Reall, {\it Dynamic dilatonic domain walls,}
hep-th/9903225; A.Chamblin, M.J. Perry and H.S.Reall, {\it Non-BPS
D8-branes and dynamic domain walls in massive IIA supergravities,
J.High Energy Phys.} {\bf{09}} {\it (1999) 014} [hep-th/9908047].

P.Kraus, {\it Dynamics of Anti-de Sitter Domain Walls, JHEP
9912:011 (1999)} [hep-th/9910149]

\bibitem{Mavro}
J. E. Ellis, N. E. Mavromatos and D. V. Nanopoulos, {\it Time
Dependent Vacuum Energy Induced by D Particle Recoil, Gen. Rel.
Grav.} {\bf{32}} {\it (2000) 943} [gr-qc/9810086]

\bibitem{Keh}
 A.Kehagias and E.Kiritsis, {\it Mirage cosmology, JHEP 9911:022 (1999)}
 [hep-th/9910174]


\bibitem{Papa}
 E.Papantonopoulos and I.Pappa, {\it Type 0 Brane Inflation from Mirage
 Cosmology, Mod. Phys. Lett.} {\bf{A 15}} {\it (2000) 2145}
  [hep-th/0001183]

 \bibitem{Kim}
 J.Y. Kim, {\it Dilaton-driven brane inflation in type IIB string
 theory}, [hep-th/0004155]

 \bibitem{Korean}
J. Y. Kim, {\it Brane inflation in tachyonic and non-tachyonic
type OB string theories} [hep-th/0009175]

\bibitem{Pappa}
 E.Papantonopoulos and I.Pappa, {\it Cosmological Evolution of a
 Brane Universe in a Type 0 String Background  Phys. Rev.}
 {\bf D63} {\it (2001) 103506 } [hep-th/0010014]

\bibitem{Youm}
D. Youm, {\it Brane Inflation in the Background of a D-Brane with
NS B field}, [hep-th/0011024]; {\it Closed Universe in Mirage
Cosmology}, [hep-th/0011290]

\bibitem{Reg}
T.Regge and C.Teitelboim,  Marcel Grossman Meeting on General
Relativity,Trieste 1975, North Holland ;

 V.A. Rubakov and M.E. Shaposhnikov, {\it Do we live inside a
 domain wall? Phys. Lett.},{\bf{B 125}} {\it (1983) 136 }.

\bibitem{Dim}
N.Arkani-Hamed, S.Dimopoulos and G.Dvali, {\it The hierarchy
problem and new dimensions at a millimeter, Phys. Lett.} {\bf{B
429}}{\it (1998) 263} [hep-ph/9803315];
{\it Phenomenology, astrophysics and cosmology of theories with
submillimeter dimensions and TeV scale quantum gravity, Phys.
Rev.}{\bf{D 59}} {\it (1999) 086004}[hep-ph/9807344];

I.Antoniadis, N.Arkani-Hamed, S.Dimopoulos and G.Dvali, {\it New
dimensions at a millimeter to a Fermi and superstrings at a TeV,
Phys. Lett.} {\bf{B 436}} {\it (1998) 257} [hep-ph/9804398]

\bibitem{Rand}
R.Sundrum, {\it Effective field theory for a three-brane universe,
Phys. Rev.} {\bf{D 59}} {\it (1999) 085009} [hep-ph/9805471];
{\it Compactification for a three brane universe, Phys. Rev.}
{\bf{D 59}} {\it (1999) 085010} [hep-ph/9807348];

L.Randall and R.Sundrum, {\it Out of this world supersymmetry
breaking, Nucl. Phys.} {\bf{B.557}} {\it (1999) 79}
[hep-th/9810155]; {\it A large mass hierarchy from a small extra
dimension, Phys. Rev. Lett. {\bf{83}}  (1999) 3370}
[hep-ph/9905221].

\bibitem{mousi}
E. Papantonopoulos and I. Pappa in preparation


\end{thebibliography}
 \end{document}